\documentclass[aps,prd,preprint,groupedaddress,showpacs]{revtex4}

\pdfoutput=1
\usepackage{amsmath}
\usepackage{amsfonts}
\usepackage{amssymb}
\usepackage{graphicx}
\usepackage{epsfig}
\usepackage{color}
\usepackage{float}
\usepackage{hyperref}
\usepackage{verbatim}
\usepackage{fancyhdr}
\usepackage{amsthm}
\usepackage{bbold}
\usepackage{verbatim}

\begin{document}

%\preprint{}

\title{Equilibration of a quantum field in de Sitter space-time}

\author{Andreas Albrecht}
\email[]{ajalbrecht@ucdavis.edu}
\affiliation{University of California at Davis, Department of Physics, One Shields Ave, Davis CA 95616 USA}
\author{R.~Holman}
\email[]{rh4a@andrew.cmu.edu}
\affiliation{Physics Department, Carnegie Mellon University, Pittsburgh PA 15213 USA}
\author{Benoit J. Richard}
\email[]{bjrichard@ucdavis.edu}
\affiliation{University of California at Davis, Department of Physics, One Shields Avenue, Davis CA 95616 USA}

\date{\today}

\begin{abstract}
We address the following question: To what extent can a quantum field
tell if it has been placed in de Sitter space? Our approach is to use
the techniques of non-equilibrium quantum field theory to compute the
time evolution of a state which starts off in flat space for
(conformal) times $\eta<\eta_0$, and then evolves in a de Sitter
background turned on instantaneously at $\eta=\eta_0$.  We find that
the answer depends on what 
quantities one examines.  We study a range of them, all based on
two-point correlation functions, and analyze which ones approach
the standard Bunch-Davies values over time. The outcome of this
analysis suggests that the nature of the equilibration process in this
system is similar to that in more familiar systems.
\end{abstract}


\maketitle

\section{Introduction\label{sec:intro}}

De Sitter space is widely accepted as a probable early-universe
cosmological solution, as it describes the state of the universe
during inflation. Provided our universe possesses a
completely stable positive cosmological constant, it should also
asymptotically approach de Sitter space at late times\textsuperscript{\footnotemark[1]}\footnotetext[1]{Such an idea is
at the heart of  de Sitter Equilibrium, an alternative to eternal
inflation as an initial conditions independent cosmological framework
\cite{AlbrechtSorbo2004, Albrecht2009}, but the motivation for this
paper is broader than this.}. 	

The standard lore of de Sitter space is that it acts as a heat bath,
in such a way that an Unruh-deWitt detector for a quantum field in a de Sitter background
will register a thermal spectrum for the number of particles in a
given momentum mode (\cite{BirrellDavies1982}
and references therein).  But how
does this happen? If de Sitter space is past and future eternal and the state is de Sitter invariant, then it should not come as a
surprise that the Green's functions of the quantum field embedded in
this background should partake of its thermal behavior as evidenced by
the periodicity in imaginary time inherent in the
metric \cite{GibbonsHawking1977}.  However, suppose we start the de
Sitter evolution at an initial time, as might happen in inflation,
say, and further assume that we start the field in a state that is not de Sitter
invariant. What happens next? 

We address this question here in a particular scheme which is chosen to
be relevant to the question and also technically tractable.  Consider the situation where, for
conformal times $\eta<\eta_0$, the background geometry is that of
Minkowski flat space-time, and a minimally coupled free field is taken
to be in the free field vacuum state of the flat space
Hamiltonian. Then at $\eta =\eta_0$, the background is changed to
become de Sitter space with an expansion rate $H$ so that the
Gibbons-Hawking temperature is  $T_{\rm de\ S} = \frac{H}{2\pi}$. Since
we are only considering free field theory in a time-dependent
background, we can solve the functional Schr\"odinger equation for the
wave functional describing the state of the field explicitly, and use
this wave functional to understand to what extent does this state
approximate the ``thermal'' Bunch-Davies (BD) state \cite{BunchDavies1978} (extending the work on the corresponding modes by Schomblond and Spindel \cite{SchomblondSpindel1976}), by analyzing ratios of various correlators and momentum-energy tensors, evaluated in our vacuum state to the quantities considered in the BD vacuum state.   

This issue is not only of conceptual relevance, but could have
observational consequences as well. The state of the field prior to
inflation need not be one that matches on smoothly to the BD state at
the onset of inflation, and if the number of e-folds is close to the
minimum it
is not an outlandish thought that some remnants of this
pre-inflationary state might have survived to imprint themselves on
the CMB and/or large scale structure. Conversely, given how well the
power spectrum of CMB fluctuations has been
measured \cite{WMAP2012,Planck2013}, and how closely this spectrum
follows what would have been expected from the assumptions of an
initial BD state, we can use this data and our calculation to
constrain the space of allowed initial states for inflationary
fluctuations.  

It is worth noting that the question we are asking can be recast as: to what extent are there no-hair theorems for the quantum state of a test field in de Sitter space? There has been some prior work in this direction, starting from the seminal work of Ford and Vilenkin \cite{VilenkinFord1982} as well as the more recent one of Anderson, Eaker, Habib, and Molina-Par\'{i}s \cite{AndersonHabibMottolaParis2000}. In both cases, an attractor behavior was found for sufficiently well-behaved states. Related issues were also addressed in \cite{FischlerKunduPedraza2013}, \cite{FischlerNguyenPedrazaTangarife2014}, \cite{SinghGangulyPadmanabhan2013}, and \cite{SinghModakPadmanabhan2013}. Our viewpoint is somewhat different here; we don't know what the state of the field is prior to inflation but, regardless, it should be reasonable to ask what the evolution of that state is after inflation begins. Then we can ask to what extent the BD behavior is generic at late times during inflation. 

In the next section we set up the initial value problem for the
Schr\"odinger wave functional with the flat space initial conditions
described above. We then use that wave functional to compute two-point
functions in our state. Since we have a free field theory, that state
will be a gaussian, and thus fully described by the three correlation functions: $\langle
\Phi_{\vec{k}}\Phi_{-\vec{k}}\rangle$, $\langle
\Pi_{\vec{k}}\Pi_{-\vec{k}}\rangle$, and $\langle \Pi_{\vec{k}}
\Phi_{-\vec{k}}+ \Phi_{\vec{k}}\Pi_{-\vec{k}}\rangle$. Additionally,
we study observables such as the
expectation value of the stress-energy tensor of this state. Section
\ref{sec:numerics} is devoted to numerical results and the analysis of
ratios of two-point functions, and stress-energy
tensors, evaluated in both our state and the BD states. We
end with a discussion of our results as well as some further
directions to take in addressing the issues dealt with here. Overall
our vacuum state approaches the BD state, when considering
coarse-grained collections of modes clearly within the horizon.

\section{\label{sec:schrodingerFT} Wave Functional, Mode Equation, and Correlation Functions}

\subsection{Finding the Schr\"odinger wave functional}

 As discussed in the introduction, we consider a test scalar
 field embedded in an FRW space-time that transitions between a
 constant scale factor for conformal times $\eta\leq \eta_0$ to a de
 Sitter scale factor for $\eta>\eta_0$. We assume that such a
 space-time is generated by an appropriate stress-energy tensor via
 the Einstein equations, but do not concern ourselves further with how
 this background geometry is obtained.   

If $\Phi(\vec{x}, \eta)$ denotes the scalar field in question, the action we use is

\begin{equation}
\label{eq:action}
S = \frac{1}{2}\int d^4 x a^4(\eta)\left[\frac{1}{a^2(\eta)} \left(\Phi^{\prime 2}-(\nabla \Phi)^2\right) -m^2 \Phi^2\right],
\end{equation}
\\
where a prime denotes a derivative with respect to $\eta$, and $m^2 = m^2_{\Phi} + \xi_BR$, $m_\Phi$ referring to the mass of our field. We will take the scale factor as 

\begin{equation}
a(\eta) = \left\{\begin{array}{cc}-\frac{1}{\eta_0 H} & \eta\leq \eta_0 \\-\frac{1}{\eta H} & \eta>\eta_0\end{array}\right.
\end{equation}
\\
The scale factor is continuous though not differentiable at
$\eta=\eta_0$. A more reasonable assumption would be that the
transition is smoother than this (for an example see
\cite{VilenkinFord1982}), but this will suffice for our purposes.  

Instead of quantizing this theory in the usual way (i.e., by defining
creation and annihilation operators acting on the Fock space of
states) we will use a Schr\"odinger picture  quantization
\cite{BoyanovskyVegaHolman1994}. Here we use eigenstates of
the  Schr\"odinger picture field operator $\hat{\Phi}(\vec{x})$,
$|\Phi(\cdot)\rangle$ such that  

\begin{equation}
\label{eq:spicquant}
\hat{\Phi}(\vec{x}) |\Phi(\cdot)\rangle=\Phi(\vec{x})  |\Phi(\cdot)\rangle.
\end{equation}
\\
The state of the field is then represented by a wave functional $\Psi\left[\Phi(\cdot); \eta\right]$ (or more generally by a density matrix element $\rho[\Phi(\cdot),\tilde{\Phi}(\cdot);\eta]$) satisfying the Schr\"odinger (Liouville) equation

\begin{equation}
\label{eq:schreqn}
i\frac{\partial \Psi[\Phi(\cdot); \eta]}{\partial \eta} = \hat{H}\left[-i\frac{\delta}{\delta \Phi(\cdot)}, \Phi(\cdot)\right] \Psi[\Phi(\cdot); \eta]\quad \left(\text{or }i\frac{\partial\rho}{\partial \eta}=\left[ \hat{H}, \rho\right]\right), 
\end{equation}
\\
where $\hat{H}$ is the Hamiltonian operator (again in the Schr\"odinger picture) obtained from the action in Eq. (\ref{eq:action}). For our case this reads

\begin{equation}
\label{eq:hamiltonian}
\hat{H} = \int d^3 x\left\{ \frac{\Pi^2}{2 a^2 (\eta)} +\frac{1}{2} a^2 (\eta) \left(\nabla \Phi(\vec{x})\right)^2 + \frac{1}{2} m^2 a^4(\eta)\Phi(\vec{x})^2 \right\},
\end{equation}
\\
with $\Pi = a^2 (\eta) \Phi^{\prime}$ being the canonically conjugate momentum to $\Phi$, represented in the usual way as $\Pi\rightarrow -i\delta\slash \delta \Phi(\cdot)$ in the Schr\"odinger picture. 

We note that the Schr\"odinger equation in
Eq. (\ref{eq:schreqn}) should  be written using the proper time of the observer
measuring the wave function. For an FRW space-time this would be the
cosmic time $t$. However, the use of conformal time corresponds to a
canonical transformation and thus gives rise to the same
physics \cite{BoyanovskyVegaHolman1994}, as would be expected of a
coordinate transformation.  It will be important in our later analysis to keep in mind that 
$t\rightarrow \infty$ corresponds to $\eta \rightarrow 0^-$.

We will take our spatial geometry to be flat so we can expand the field in terms of Fourier components. Furthermore, we will quantize our field in a box of comoving spatial volume $V$ so that the Schr\"odinger picture field and conjugate momenta can be written as

\begin{eqnarray}
\label{eq:expansion}
& & \Phi(\vec{x}) = \frac{1}{\sqrt{V}} \sum_{\vec{k}} \Phi_{\vec{k}} e^{-i \vec{k}\cdot \vec{x}}\nonumber\\
& & \Pi(\vec{x}) =  \frac{1}{\sqrt{V}} \sum_{\vec{k}} \Pi_{\vec{k}} e^{-i \vec{k}\cdot \vec{x}},
\end{eqnarray}
\\
where the equal time commutation relations 

\begin{equation}
\label{eq:commrel}
\left[ \Phi_S(\vec{x}),\Pi_S(\vec{y})\right]= i\delta^3(\vec{x}-\vec{y})
\end{equation}
 \\
imply $\left[\Phi_{\vec{k}}, \Pi_{\vec{q}}\right] = i \delta_{\vec{k}, -\vec{q}}$ and thus, in the Schr\"{o}dinger picture, $\Pi_{\vec{q}}$ can be represented as $-i\frac{\delta}{\delta \Phi_{\vec{-q}}}$. Hence, the Hamiltonian breaks up into the sum of Hamiltonians for each mode, and we can also write the wave function as the product of wave functions for each mode: 

\begin{eqnarray}
\label{eq:momentumspace}
&& H = \sum_{\vec{k}} H_{\vec{k}}\quad  \text{with } H_{\vec{k}}= \frac{\Pi_{\vec{k}} \Pi_{-\vec{k}}}{2 a^2(\eta)} +\frac{1}{2} a^2(\eta)  \Omega_{\vec{k}}^2(\eta)\ \Phi_{\vec{k}} \Phi_{-\vec{k}},\nonumber\\
&& \Psi[\{\Phi_{\vec{k}}\}, \eta] = \prod_{\vec{k}} \psi_{\vec{k}} (\Phi_{\vec{k}}, \eta),\nonumber\\
&&  \Omega_k^2(\eta) \equiv k^2 + m^2 a^2(\eta).
\end{eqnarray}
\\

Since we have a free field theory, our ansatz for the ground-state wave functional for each mode should be Gaussian as in

\begin{equation}
\label{eq:gaussianansatz}
\psi_{\vec{k}} (\Phi_{\vec{k}}, \eta)=N_{\vec{k}} (\eta) \exp\left(-\frac{1}{2} A_k (\eta)  \Phi_{\vec{k}} \Phi_{-\vec{k}}\right),
\end{equation}
\\
where we have made use of rotational invariance to write the kernel $A_k(\eta)$ as a function of the magnitude $k$ of $\vec{k}$. By matching powers of $\Phi_{\vec{k}}$ on either side of the Schr\"odinger equation for each mode we find:

\begin{eqnarray}
\label{eq:seqn}
&& i\frac{N_{\vec{k}}^{\prime} (\eta)}{N_{\vec{k}} (\eta)} = \frac{A_k(\eta)}{2 a^2(\eta)} \nonumber\\
&& i A_k^{\prime}(\eta) = \frac{A_k^2(\eta)}{a^2(\eta)}-a^2(\eta)\Omega_{k}^2(\eta)\quad A_k(\eta_0) = \Omega_k(\eta_0) a^2(\eta_0)
\end{eqnarray}
\\
where the primes represent conformal time derivatives, and the initial condition is found by considering the ground state wave function of a quantum mechanical harmonic oscillator with mass $a^2(\eta_0)$ and frequency $\Omega_k (\eta_0)$.

\subsection{Solving the mode equations}

 Eq. \eqref{eq:seqn} is of the Ricatti form and can be converted into a second order equation of Schr\"odinger type via the substitution 

\begin{equation}
A_k(\eta) = -i a^2(\eta)\left(\frac{\phi_k^{\prime}(\eta)}{\phi_k(\eta)}-\frac{a^{\prime}(\eta)}{a(\eta)}\right).
\end{equation}
\\
Doing this we find

\begin{equation}
\label{eq:mode}
\phi_k^{\prime \prime}(\eta) + \left(\Omega_k^2(\eta)-\frac{a^{\prime \prime}(\eta)}{a(\eta)}\right) \phi_k(\eta)=0,\quad \phi_k^{\prime}(\eta_0) =\left( i  \Omega_k(\eta_0)+\frac{a^{\prime}(\eta_0)}{a(\eta_0)}\right)\phi_k(\eta_0).
\end{equation}
\\
The equation we start with for $A_k(\eta)$ is a first order equation and we have one initial condition for it so that there is a unique solution for $A_k(\eta)$. On the other hand, the equation for $\phi_k(\eta)$ is a second order one, requiring two initial conditions for a unique solution. The resolution of this dilemma can be found by noting that $A_k$ is related to the ratio of $\phi_k^{\prime}$ and $\phi_k$. This means that in any linear combination of the two independent solutions to Eq. (\ref{eq:mode}), we can factor out an overall constant leaving only one constant to be determined. We can use this freedom to fix the (constant) Wronskian of $\phi_k(\eta)$ and  $\phi_k^*(\eta)$ to equal $-i$. Imposing this condition then implies that $\phi_k(\eta_0) = \frac{1}{\sqrt{2 \Omega_k(\eta_0)}}$.

Eq. (\ref{eq:mode}) is nothing but the mode equation for a massive, minimally coupled scalar field in de Sitter space. The solutions are well known \cite{SchomblondSpindel1976} and we can write

\begin{eqnarray}
\label{eq:modesoln}
&& \phi_k(\eta) = \alpha_k {\cal U}_k(\eta)+\beta_k  {\cal U}_k^*(\eta),\quad {\cal U}_k(\eta) = \frac{\sqrt{-\pi \eta}}{2}H_{\nu}^{(2)}(-k\eta),\nonumber\\
&& \alpha_k = \frac{i}{\sqrt{2 \Omega_k(\eta_0)}}\left[{\cal U}_k^{* \prime}(\eta_0)+ \left(-i \Omega_k(\eta_0)+\frac{1}{\eta_0}\right){\cal U}_k^{*}(\eta_0)\right], \\ 
&&\beta_k = -\frac{i}{\sqrt{2 \Omega_k(\eta_0)}}\left[{\cal U}_k^{ \prime}(\eta_0)+\left(-i \Omega_k(\eta_0)+\frac{1}{\eta_0}\right){\cal U}_k(\eta_0)\right],\nonumber
\end{eqnarray}
\\
where $\nu =
\sqrt{\frac{9}{4}-\frac{m^2}{H^2}}$ and ${\cal U}_k(\eta)$ is commonly referred to as the $k^{th}$ Bunch-Davies mode. It is easy to check that the
Wronskian condition implies that $|\alpha_k|^2-|\beta_k|^2=1$; had we
been doing Heisenberg field theory, we would infer that the modes
$\phi_k(\eta)$ are just the Bogoliubov transforms of the BD modes. Moreover, as $\eta_0 \rightarrow -\infty$, the
form of ${\cal U}_k(\eta)$ allows us to conclude:

\begin{eqnarray}
\label{eq:limeta0}
&& \Omega_k(\eta_0) \rightarrow k \nonumber \\
&&{\cal U}_k(\eta_0) \rightarrow \frac{1}{\sqrt{2k}}, \\
&&{\cal U'}_k(\eta_0) \rightarrow i \sqrt{\frac{k}{2}} \nonumber,
\end{eqnarray}
\\
from which we can infer $\alpha_k \rightarrow 1$ and $\beta_k \rightarrow 0$, i.e. in this limit, we go back to an eternal inflationary patch of de Sitter space with the field state being the BD state.

The full wave function for the mode $\Phi_{\vec{k}}$ is thus given by

\begin{equation}
\label{eq:kwavefcn}
\psi_{\vec{k}}(\eta) = \left(\frac{a^2(\eta)}{\pi \left|\phi_k(\eta)\right|^2}\right)^{\frac{1}{4}} \exp\left[\frac{i}{2} a^2(\eta) \left(\frac{\phi_k^{\prime}(\eta)}{\phi_k(\eta)}-\frac{a^{\prime}(\eta)}{a(\eta)}\right) \Phi_{\vec{k}} \Phi_{-\vec{k}}\right],
\end{equation}
\\
where we should note that when computing any expectation values for
quantities involving the mode $\Phi_{\vec{k}}$, we also need to
include the contribution of the wave function $\psi_{-\vec{k}}(\eta)$,
since $\Phi_{-\vec{k}} = \Phi_{\vec{k}}^*$, and $\Phi$ is a real
field. This is equivalent to using the square of
$\psi_{\vec{k}}(\eta)$ in any such calculation.  

Eq. \eqref{eq:kwavefcn} coupled with the mode equations (\ref{eq:mode}) gives the full specification of the quantum state with the given initial conditions. We can now use this wave function to compute observables that might help us answer the question asked in the introduction: to what extent does this state ``feel'' de Sitter space?

\subsection{Calculating relevant correlation functions}

 What are the useful diagnostic tools to evaluate the behavior of this state? Since the state is Gaussian, it can be fully specified by the following correlators: $\langle \Phi_{\vec{k}}\Phi_{-\vec{k}}\rangle,\ \langle \Pi_{\vec{k}}\Pi_{-\vec{k}}\rangle,\  \langle \Pi_{\vec{k}} \Phi_{-\vec{k}}+ \Phi_{\vec{k}}\Pi_{-\vec{k}}\rangle$, computed below. 

From (\ref{eq:kwavefcn}) computing $\langle \Phi_{\vec{k}} \Phi_{-\vec{k}}\rangle(\eta)$ results in

\begin{eqnarray}
\label{eq:2ptphi}
\langle \Phi_{\vec{k}} \Phi_{-\vec{k}}\rangle(\eta) &&= \int {\cal D}\Phi_{\vec{k}} \left|\psi_{\vec{k}}(\eta)\right|^2  \left|\psi_{-\vec{k}}(\eta)\right|^2 \Phi_{\vec{k}} \Phi_{-\vec{k}} \nonumber \\
&& = \frac{1}{2A_{kR}} \\
&& = \frac{|\phi_k(\eta)|^2}{a^2(\eta)}. \nonumber
\end{eqnarray}
\\
where $A_{kR}$ denotes the real part of the kernel $A_k(\eta)$.

The other correlators are also easy enough to compute. For $\langle \Pi_{\vec{k}}\Pi_{-\vec{k}}\rangle$ we have

\begin{eqnarray}
\label{eq:2ptpi}
\langle \Pi_{\vec{k}} \Pi_{-\vec{k}}\rangle(\eta) &&= \int {\cal D}\Phi_{\vec{k}}\ \psi_{\vec{k}}(\eta)^{* 2}  \left(-\frac{\delta^2}{\delta \Phi_{\vec{k}} \delta \Phi_{-\vec{k}}}\right) \psi_{-\vec{k}}(\eta)^2 \nonumber \\
&& =\frac{\left|A_k\right|^2}{2 A_{k R}} \\
&& = a^4(\eta) \left| \frac{d}{d\eta}\left(\frac{\phi_k(\eta)}{a}\right) \right|^2. \nonumber
\end{eqnarray}
\\
 Finally, we find $\langle \Pi_{\vec{k}} \Phi_{-\vec{k}}+ \Phi_{\vec{k}}\Pi_{-\vec{k}}\rangle$ to be given by

\begin{eqnarray}
\label{eq:2ptmixed}
\langle \Pi_{\vec{k}} \Phi_{-\vec{k}}+ \Phi_{\vec{k}}\Pi_{-\vec{k}}\rangle&& =\int {\cal D}\Phi_{\vec{k}}\ \psi_{\vec{k}}(\eta)^{* 2} \left(-i\frac{\delta}{\delta \Phi_{-\vec{k}}}\Phi_{\vec{k}} -\Phi_{-\vec{k}}i\frac{\delta}{\delta \Phi_{\vec{k}}} \right) \psi_{-\vec{k}}(\eta)^2 \nonumber \\
&& = -\frac{A_{k I}}{A_{k R}}\\
&&=a^2(\eta) \frac{d}{d\eta}\left(\frac{\left|\phi_k(\eta)\right|^2}{a^2(\eta)}\right) \nonumber.
\end{eqnarray}
\\
We can check to see what happens to our two-point functions as a
function of time. In particular, we might expect that, if de Sitter
space really did act as a heat bath and an ``equilibration'' process truly 
was in effect over time, then we should see these correlators approach
the standard de Sitter two-point functions. We can check this by
noticing that at late times, it is only the imaginary part of the
Hankel function that becomes relevant (since it is singular as
$\eta\rightarrow 0^-$). Hence, the late-time expression of the Hankel
function is 

\begin{equation*}\lim_{\eta \to 0^-} H_{\nu}^{(2)}(k\eta) = i\frac{\Gamma(\nu)}{\pi}\left(\frac{2}{-k\eta}\right)^{\nu}, \end{equation*}
\\
so that, as $k\eta$ approaches $0^-$ for finite $k$, we can use this form in our two-point functions to find:

\begin{align}
\label{eq:2ptfunctionsAsymptotic}
&\left\langle \Phi_{\vec{k}}\Phi_{-\vec{k}} \right\rangle \rightarrow 4^{\nu -1}\frac{H^2  \Gamma^2(\nu)}{\pi}\left|\alpha_k - \beta_k \right|^2(-\eta)^{3} (-k\eta)^{-2\nu} , \nonumber \\ 
& \left\langle\Pi_{\vec{k}}\Pi_{-\vec{k}} \right\rangle \rightarrow 4^{\nu - 3}\frac{\left|\alpha_k - \beta_k \right|^2}{\pi H^2} (-\eta)^{-3}(-k\eta)^{-2\nu}\left[(k\eta)^2\Gamma(\nu-1)+(6-4 \nu)\Gamma(\nu)\right]^2, \\
& \left\langle \Phi_{\vec{k}}\Pi_{-\vec{k}} +\Pi_{-\vec{k}}\Phi_{\vec{k}} \right\rangle \rightarrow -4^{\nu-\frac{3}{2}} \frac{\left|\alpha_k - \beta_k \right|^2}{\pi} (-k\eta)^{-2\nu}\Gamma(\nu)\left[(k\eta)^2\Gamma(\nu-1)+(6-4\nu)\Gamma(\nu)\right]. \nonumber
\end{align} 
\\
From \eqref{eq:modesoln}, we can compute $ \left|\alpha_k-\beta_k\right|^2$ as

\begin{equation}
 \left|\alpha_k-\beta_k\right|^2=\frac{2}{\Omega_k(\eta_0)}\left[\left(\Re\left({\cal
     U}_k^{\prime}(\eta_0)+\frac{1}{\eta_0}{\cal
     U}_k(\eta_0)\right)\right)^2+\Omega_k^2(\eta_0)
   \left(\Re\left({\cal U}_k(\eta_0)\right)\right)^2\right].
\label{EqnCoeff}
\end{equation}
\\
At first glance, focusing our attention on $\left\langle
\Phi_{\vec{k}}\Phi_{-\vec{k}} \right\rangle$, Eq. 
(\ref{eq:2ptfunctionsAsymptotic}) tells us that, even at late times,
information about the initial state as encoded in the coefficients
$\alpha_k$ and $ \beta_k$ is not lost, at least not in the two-point
function. This should not be surprising since unitary evolution
always preserves information about the initial state as long as the
state is viewed in a sufficiently fine-grained manner.  It is only
through coarse-graining that a process of equilibration (should it
occur) will be revealed.  In this paper we will consider coarse-graining that is expressed by looking at quantities averaged over a
range of $k$ modes.  

We can be more explicit about Eq. \eqref{EqnCoeff} in the massless, minimally
coupled case where $\nu= \frac{3}{2}$ (this was also treated
in \cite{AndersonHabibMottolaParis2000}). In this case,  

\begin{equation}
{\cal U}_k(\eta) = -\frac{e^{ik\eta}}{\sqrt{2 k}}\left(1+\frac{i}{k \eta}\right),\quad \Omega_k(\eta_0) = k,
\end{equation}
\\
and 

\begin{equation}
\label{eq:bogomass}
\left|\alpha_k-\beta_k\right|^2 = 1-\frac{\sin 2 k \eta_0}{k \eta_0} + \frac{\sin^2 k \eta_0}{\left(k \eta_0\right)^2}.
\end{equation}
\\
For $-k\eta_0\gg 1$, this modulating factor tends towards $1$. 

The two-point functions are studied in further detail in section~\ref{sec:numerics}, after substituting $q = -k\eta_0$ in the mode equation, and performing proper rescalings of our quantities by appropriate powers of $-\eta_0$.

\subsection{The stress-energy tensor}

With the previous two-point functions in hand, we can study the equilibration of our mode with respect to the BD mode. Moreover, we can calculate relevant quantities such as the stress-energy tensor and in particular the energy density $\left\langle T^0_{\phantom{b}0} \right\rangle$.

We need to compute the expectation value of the stress-energy tensor in a particular state corresponding to the density matrix $\rho(\eta)$. The momentum-energy tensor in operator form is

\begin{align*} T_{\mu\nu} = &\left(1-2\xi_B\right)\nabla_\mu\Phi\nabla_\nu\Phi + \left(2\xi_B-\frac{1}{2}\right)g_{\mu\nu}g^{\alpha\beta}\nabla_\alpha\nabla_\beta\Phi + g_{\mu\nu} V(\Phi) \\
&-\xi_B\Phi^2\left(R_{\mu\nu} - \frac{1}{2}g_{\mu\nu}R\right) + 2\xi_B\Phi(g_{\mu\nu}\Box - \nabla_\mu\nabla_\nu)\Phi.
\end{align*}
\\
Thus, in a de Sitter background,

\begin{align}
\label{eq:T00}
%\begin{split}
\left\langle T^0_{\phantom{b}0} \right\rangle = & \left\langle \frac{\Phi'^2}{2a^2} + \frac{1}{2a^2}(1-4\xi_B)(\nabla \Phi)^2 + V(\Phi) - \xi_B G^0_{\phantom{b}0}\Phi^2  + 2 \xi_B \Phi \left[3\frac{a'}{a^3}\Phi' - \frac{1}{a^2} \nabla^2\Phi \right] \right\rangle,
%\end{split}
\end{align}
\\
 where $G^{\mu}_{\phantom{b}\nu} = R^{\mu}_{\phantom{b}\nu} - \frac{1}{2} \delta^{\mu}_{\phantom{b}\nu} R$ is the Einstein tensor. $\rho(\eta)$ is written in terms of the Fourier components of the fluctuation fields. Hence, $T^0_{\phantom{b}0}$ also needs to be expanded in terms of such fluctuations. Additionally, we have $\pi = \frac{\Phi'}{a^2}$ and need to hermitianize $\Phi\Phi'$ so that $\Phi\Phi'$ goes to $\frac{1}{2a^2}(\Phi\tilde{\pi} + \tilde{\pi}\Phi)$. Therefore, combining the previous results with Eq. \eqref{eq:T00}, one obtains

\begin{align}
\label{eq:T00Fourier}
\left\langle T^0_{\phantom{b}0} \right\rangle = \int \frac{d^3k}{(2\pi)^3} & \Bigg[ \frac{1}{2a^6} \left\langle\Pi_{\vec{k}}\Pi_{-\vec{k}}\right\rangle + \left(\frac{1}{2a^2}(k^2 + a^2 V''(\Phi)) - \xi_BG^0_{\phantom{b}0} \right) \left\langle \Phi_{\vec{k}}\Phi_{-\vec{k}}\right\rangle \nonumber \\
&+3\xi_B\frac{a'}{a^5}\left\langle \Phi_{\vec{k}}\Pi_{-\vec{k}} + \Pi_{\vec{k}}\Phi_{-\vec{k}}\right\rangle \Bigg],
\end{align}
\\
where in our case, $a = -\frac{1}{\eta H}$, $V''(\Phi)= m^2, \frac{m^2}{H^2} = \frac{9}{4} - \nu^2$, and  $G^0_{\phantom{b}0} = -3H^2$. Because of the divergences notably appearing in  $\left\langle\Pi_{\vec{k}}\Pi_{-\vec{k}}\right\rangle$ and $\left\langle \Phi_{\vec{k}}\Pi_{-\vec{k}} + \Pi_{\vec{k}}\Phi_{-\vec{k}}\right\rangle$, the previous integration is not straight forward, even in the massless minimally coupled case. A more in depth analysis of  Eq. \eqref{eq:T00Fourier} will be presented in the next section.

\section{Numerical Work}\label{sec:numerics}

\subsection{Numerical approach}

Now that we have calculated all the relevant correlation functions involving our state and used those to compute the pertinent observables, we turn to a numerical analysis of these quantities.

Before doing this, however,  a rescaling of our mode
equations, fields $\phi_k$ and corresponding momenta $\Pi_k$ should be
performed. Time will be measured in units of $\eta_0$, $\langle
\Phi_{\vec{k}}\Phi_{-\vec{k}}\rangle$  in units of $(\eta_0H)^2$ and
$\langle \Pi_{\vec{k}}\Pi_{-\vec{k}}\rangle$ in units of
$(\eta_0H)^{-2}$. Rescale the momenta by $\eta_0$ and the modes by $\sqrt{-\eta_0}$, and let $q = -k\eta_0$ and let $x = 1 - \frac{\eta}{\eta_0}$ represent our new ``time'' variable. Since we are only interested in conformal times $\eta \in [\eta_0,0)$, we have $x \in [0,1)$. A given mode labeled by the comoving wavenumber $k$ crosses the de Sitter horizon when $k\eta= -1$, which corresponds to $x = 1 - \frac{1}{q}$.

 Then the BD mode and mode equations \eqref{eq:mode} respectively become
 
\begin{equation}
\label{eq:BDresc}
{\cal U}_q(x) = \frac{\sqrt{\pi(1-x)}}{2}H^{(2)}_\nu\left(q(1-x)\right), 
\end{equation}
\\
and

\begin{equation}
\label{eq:moderesc}
\phi_q^{\prime \prime}(x) + \left(q^2+\frac{\frac{1}{4}-\nu^2}{(1-x)^2}\right) \phi_q(x)=0,
\end{equation} 
\\
 where a prime now denotes a derivative with respect to $x$. Notice that, going from Eq. \eqref{eq:mode} to Eq. \eqref{eq:moderesc}, we made the substitution $\frac{m^2}{H^2} = \frac{9}{4} - \nu^2$. The initial conditions previously defined now give

\begin{equation}
\label{eq:ICresc}
\phi_q(0) = \frac{1}{\sqrt{2}\left(q^2+\frac{9}{4}-\nu^2\right)^{\frac{1}{4}}} \text{\hspace{10 pt} and \hspace{10 pt}} \phi_q^{\prime}(0) = \left[i\left(q^2 + \frac{9}{4} - \nu^2\right)^{\frac{1}{2}}+1\right]\phi_q(0).
\end{equation}
\\ 
The measure of the equilibration of our state to the BD state will be quantified by the approach of the corresponding correlators to the standard BD ones. We will examine this approach both mode by mode as well as in terms of momentum integrated quantities. For simplicity, we focus on the massless, minimally coupled case below.

\subsection{Correlators}\label{subsec:correlators}
\subsubsection{Single mode case}
We consider ratios of the form
$\frac{\left\langle\Phi_{\vec{q}}\Phi_{-\vec{q}}
  \right\rangle^{(Mode)}}{\left\langle\Phi_{\vec{q}}\Phi_{-\vec{q}}
  \right\rangle^{(BD)}}$,
$\frac{\left\langle\Pi_{\vec{q}}\Pi_{-\vec{q}}
  \right\rangle^{(Mode)}}{\left\langle\Pi_{\vec{q}}\Pi_{-\vec{q}}
  \right\rangle^{(BD)}}$,and $\frac{\left\langle
  \Phi_{\vec{q}}\Pi_{-\vec{q}} +\Pi_{\vec{q}}\Phi_{-\vec{q}}
  \right\rangle^{(Mode)}}{ \left\langle \Phi_{\vec{q}}\Pi_{-\vec{q}}
  +\Pi_{\vec{q}}\Phi_{-\vec{q}} \right\rangle^{(BD)}}$, where $(Mode)$
stands for a correlation function evaluated in our ansatz and $(BD)$
for the same quantity examined in the Bunch-Davies state. Below are
plots of all such ratios.

\begin{figure}[htbp]
	\centering
	\includegraphics[width=5 in, keepaspectratio]{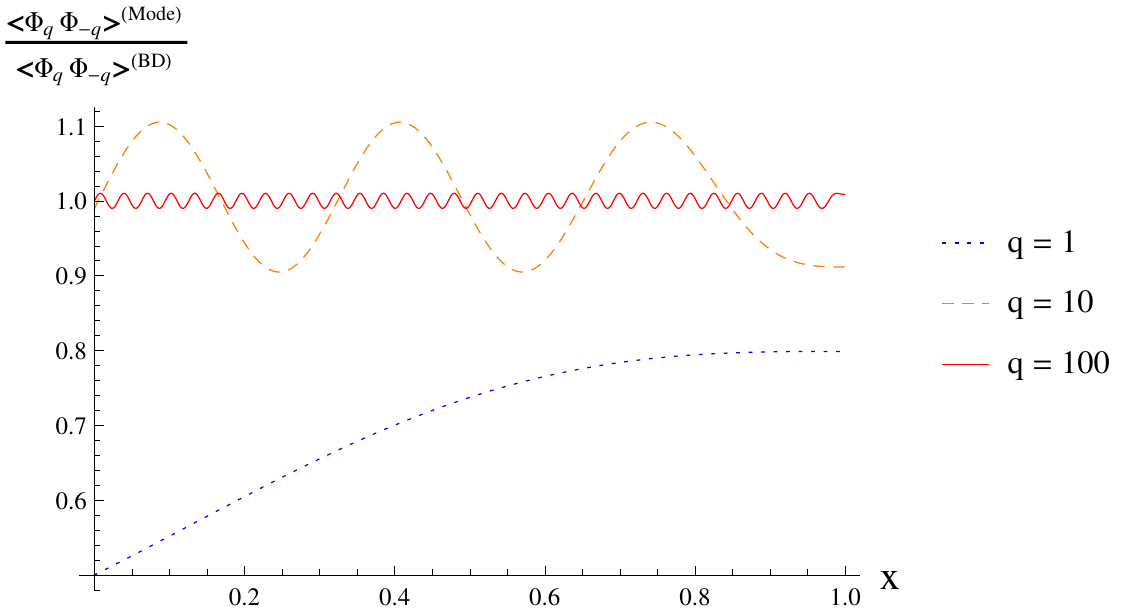}
	\caption{The ratio$\frac{\left\langle\Phi_{\vec{q}}\Phi_{-\vec{q}}
            \right\rangle^{(Mode)}}{\left\langle\Phi_{\vec{q}}\Phi_{-\vec{q}}
            \right\rangle^{(BD)}}$ for $q = 1, 10,$ and $100$. For $q
          = 1$ (dotted line) the ratio clearly does not asymptote to
          $1$, while for $q = 10$ and $q = 100$ (dashed and solid
          lines, respectively) it oscillates about $1$ without any
          damping, but with amplitude decreasing with increasing $q$
          values. Such fine-grained curves do not observe equilibration.} 
	\label{phiRatio3q}
\end{figure}

Fig. \ref{phiRatio3q} shows
$\frac{\left\langle\Phi_{\vec{q}}\Phi_{-\vec{q}}
  \right\rangle^{(Mode)}}{\left\langle\Phi_{\vec{q}}\Phi_{-\vec{q}}
  \right\rangle^{(BD)}}$ for $q = 1, 10$, and 100. For $q  = 1$,
corresponding to a mode that is crossing the horizon at $\eta = \eta_0$, the ratio seems to settle well below unity, increasing
monotonically until it plateaus for larger $x$ values, meaning that no
equilibrium between $\left\langle\Phi_{\vec{q}}\Phi_{-\vec{q}}
\right\rangle^{(Mode)}$ and $\left\langle\Phi_{\vec{q}}\Phi_{-\vec{q}}
\right\rangle^{(BD)}$ is reached. This is not
surprising. Indeed, since all modes for which $q \in [0,1]$ are
essentially frozen we should not expect anything
dynamical to happen to their matching correlation functions. Thus
it seems clear that for modes crossing the horizon or outside of it,
information about the initial state is never lost.  

As $q$ increases to 10 or even 100,
$\frac{\left\langle\Phi_{\vec{q}}\Phi_{-\vec{q}}
  \right\rangle^{(Mode)}}{\left\langle\Phi_{\vec{q}}\Phi_{-\vec{q}}
  \right\rangle^{(BD)}}$ is characterized by an undamped oscillatory
behavior about $1$ with higher $q$'s having  smaller  amplitudes. The absence of damping is  due to the fact that
taking a ratio of $\left\langle\Phi_{\vec{q}}\Phi_{-\vec{q}}
\right\rangle$ in different states erases the contributions of the
scale factors, as can be seen from Eq. \eqref{eq:2ptphi}, hence the
red-shifting of the modes due to the expansion of the
universe is removed. Additionally, since our state can be viewed as a Bogoliubov
transform of the BD state, $|\phi_q(x)|^2$ just oscillates about
$|{\cal U}_q(x)|^2$ with constant amplitude. Given the form of
$\left\langle\Phi_{\vec{q}}\Phi_{-\vec{q}} \right\rangle$ the same
should occur between $\left\langle\Phi_{\vec{q}}\Phi_{-\vec{q}}
\right\rangle^{(Mode)}$ and $\left\langle\Phi_{\vec{q}}\Phi_{-\vec{q}}
\right\rangle^{(BD)}$. 

\begin{figure}[htbp]
	\centering
	\includegraphics[width=5 in, keepaspectratio]{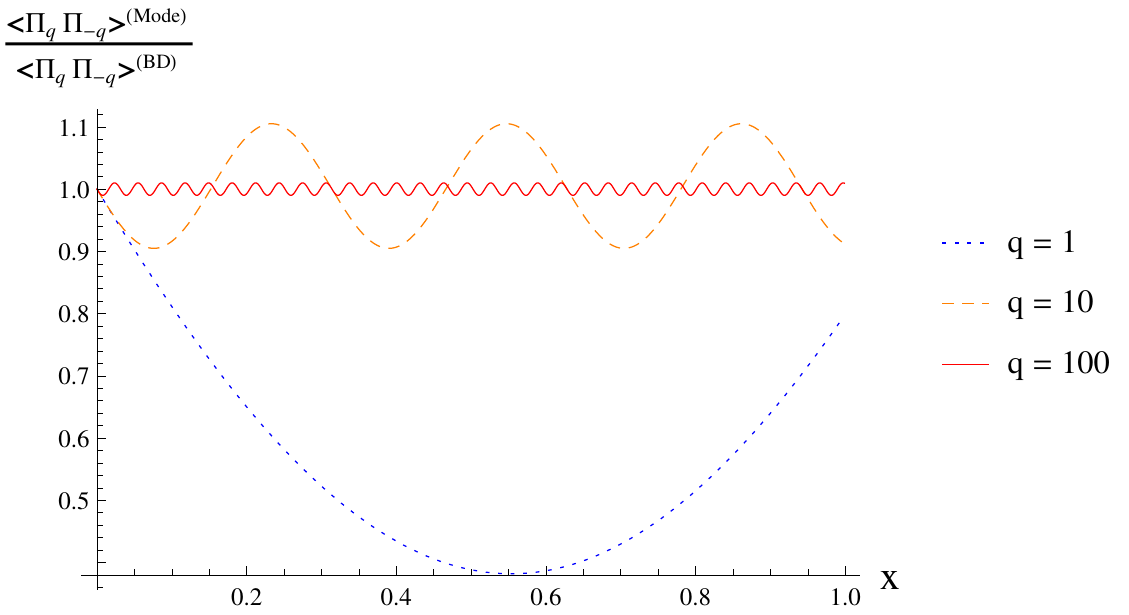}
	\caption{The ratio $\frac{\left\langle\Pi_{\vec{q}}\Pi_{-\vec{q}} \right\rangle^{(Mode)}}{\left\langle\Pi_{\vec{q}}\Pi_{-\vec{q}} \right\rangle^{(BD)}}$ for  $q = 1, 10$, and 100. Similarly to what was observed in Fig. \ref{phiRatio3q}, it appears the ratio evaluated at $q = 1$ does not approach $1$, while $\frac{\left\langle\Pi_{\vec{q}}\Pi_{-\vec{q}} \right\rangle^{(Mode)}}{\left\langle\Pi_{\vec{q}}\Pi_{-\vec{q}} \right\rangle^{(BD)}}$ taken for $q = 10$ or $q = 100$ oscillates without damping about $1$, with an amplitude that decreases as $q$ becomes higher. As in Fig. \ref{phiRatio3q} no signs of equilibration are found.}
	\label{piRatio3q}
\end{figure}

The ratio $\frac{\left\langle\Pi_{\vec{q}}\Pi_{-\vec{q}}
  \right\rangle^{(Mode)}}{\left\langle\Pi_{\vec{q}}\Pi_{-\vec{q}}
  \right\rangle^{(BD)}}$ in Fig. \ref{piRatio3q}, presents
similarities with Fig. \ref{phiRatio3q} for $q = 10$ and $q = 100$,
namely the ratio corresponding to such modes is oscillatory about an
equilibrium value of $1$ and undamped. Moreover for $q = 1$, no approach to unity is observed.

\begin{figure}[htbp] 
	\centering
	\includegraphics[width=5 in, keepaspectratio]{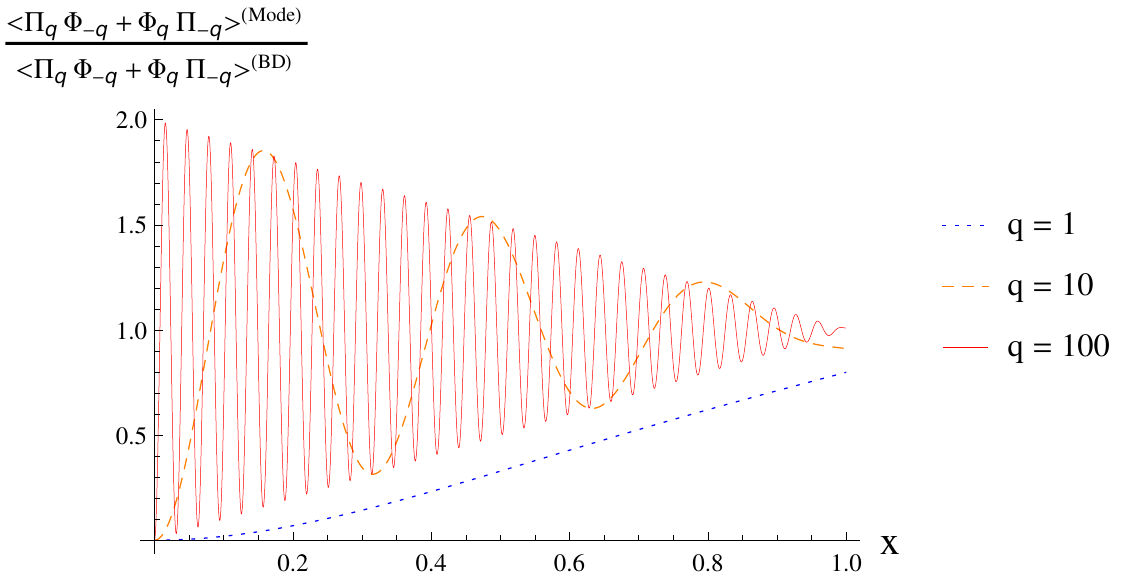}
	\caption{The ratio $\frac{\left\langle  \Pi_{\vec{q}}\Phi_{-\vec{q}} + \Phi_{\vec{q}}\Pi_{-\vec{q}} \right\rangle^{(Mode)}}{ \left\langle  \Pi_{\vec{q}}\Phi_{-\vec{q}} + \Phi_{\vec{q}}\Pi_{-\vec{q}} \right\rangle^{(BD)}}$ for  $q = 1, 10$, and $100$. Similar conclusions to the ones in Fig. \ref{phiRatio3q} and \ref{piRatio3q} can be made, namely, the higher the $q$-mode the smaller the amplitude of oscillations. However, contrary to the previous two figures, the ratios corresponding to $q = 10$ and $100$ oscillate about $1$ with amplitude decreasing linearly.}
	\label{mixRatio3q}
\end{figure}

The last ratio of correlators, $\frac{\left\langle
  \Phi_{\vec{q}}\Pi_{-\vec{q}} +\Pi_{\vec{q}}\Phi_{-\vec{q}}
  \right\rangle^{(Mode)}}{ \left\langle \Phi_{\vec{q}}\Pi_{-\vec{q}}
  +\Pi_{\vec{q}}\Phi_{-\vec{q}} \right\rangle^{(BD)}}$, is shown
in Fig. \ref{mixRatio3q}. When $q = 1$, this
quantity appears to grow monotonically for all $x$, corroborating the
absence of equilibrium for such corresponding modes.  The striking
feature of Fig. \ref{mixRatio3q}, manifesting itself when compared
to Fig. \ref{phiRatio3q} and \ref{piRatio3q}, is that the ratios
tend to oscillate about $1$ for $q > 1$, but now with an amplitude
that diminishes as a function of time. The size of the oscillations is
now damped linearly, while being constant in the first two plots. Such
a disparity is due to the presence of scale factors in
\eqref{eq:2ptmixed} that do not cancel upon taking a quotient of
two-point functions evaluated in different states.  

%used to be H

Despite the described differences found when comparing Fig.
\ref{phiRatio3q}, \ref{piRatio3q}, and \ref{mixRatio3q}, we argue that
one can draw similar conclusions regarding the lack of
equilibration. It is not surprising that 
correlation functions for modes crossing, near crossing, or outside
the horizon do not exhibit equilibration, as such modes freeze
out. Moreover the oscillatory behavior shown for higher $q$ modes in
Fig. \ref{phiRatio3q} and \ref{piRatio3q} also does not reflect
equilibration.  At first glance, the curves in Fig.~\ref{mixRatio3q}
seem to indicate an approach to the BD mode, since all curves approach unity over
time.  However, $x \rightarrow 1$ corresponds to $t \rightarrow
+\infty$ for cosmic time $t$.  We feel the slowness of the approach to
unity of the curves in Fig.~\ref{mixRatio3q} leaves us unconvinced
that this quantity should be regarded as equilibrating.  The $q=1$ and
$q=10$ curves clearly do not actually reach unity as $x \rightarrow 1$, and 
the same is true of the $q=100$ curve, although this is harder to see
from the plot. 

The lack of equilibration of the two-point functions for single modes is
hardly the final word.  After all one cannot learn about the
equilibration of a box of gas by following a single energy eigenstate
of the microscopic system,
no matter how strongly the equilibration is realized overall. We next
consider correlation functions averaged over a range 
of $q$'s, as a way to represent coarse-graining.  Although our setup
is rather formal, we believe these averaged quantities bring us closer
to representing realistic observables. 

\subsubsection{Quantities averaged over modes}
 
We integrated all our two-point functions over finite
ranges of $q$: $[1,3]$, $[3,9]$, and $[10, 20]$. Such domains in $q$
have been chosen to demonstrate the difference between modes that sit
inside the horizon (with large wavelengths for $q \in [3, 9]$ or $q$
an order of magnitude away from horizon-crossing for $q \in [10, 30]$)
and those that are traversing or near the horizon.
We have found that
the general behaviors can be identified without including even
higher values of $q$.  Our ratios then become
$\frac{\left\langle\Phi_{\vec{q}}\Phi_{-\vec{q}}
  \right\rangle^{(Mode)}_{[q_{min},
      q_{max}]}}{\left\langle\Phi_{\vec{q}}\Phi_{-\vec{q}}
  \right\rangle^{(BD)}_{[q_{min}, q_{max}]}}$,
$\frac{\left\langle\Pi_{\vec{q}}\Pi_{-\vec{q}}
  \right\rangle^{(Mode)}_{[q_{min},
      q_{max}]}}{\left\langle\Pi_{\vec{q}}\Pi_{-\vec{q}}
  \right\rangle^{(BD)}_{[q_{min}, q_{max}]}}$,and $\frac{\left\langle
  \Phi_{\vec{q}}\Pi_{-\vec{q}} +\Pi_{\vec{q}}\Phi_{-\vec{q}}
  \right\rangle^{(Mode)}_{[q_{min}, q_{max}]}}{ \left\langle
  \Phi_{\vec{q}}\Pi_{-\vec{q}} +\Pi_{\vec{q}}\Phi_{-\vec{q}}
  \right\rangle^{(BD)}_{[q_{min}, q_{max}]}},$ where $q_{min}$ is our
lower limit of integration and $q_{max}$ our upper limit. 

\begin{figure}[htbp]
	\centering
	\includegraphics[width=5 in, keepaspectratio]{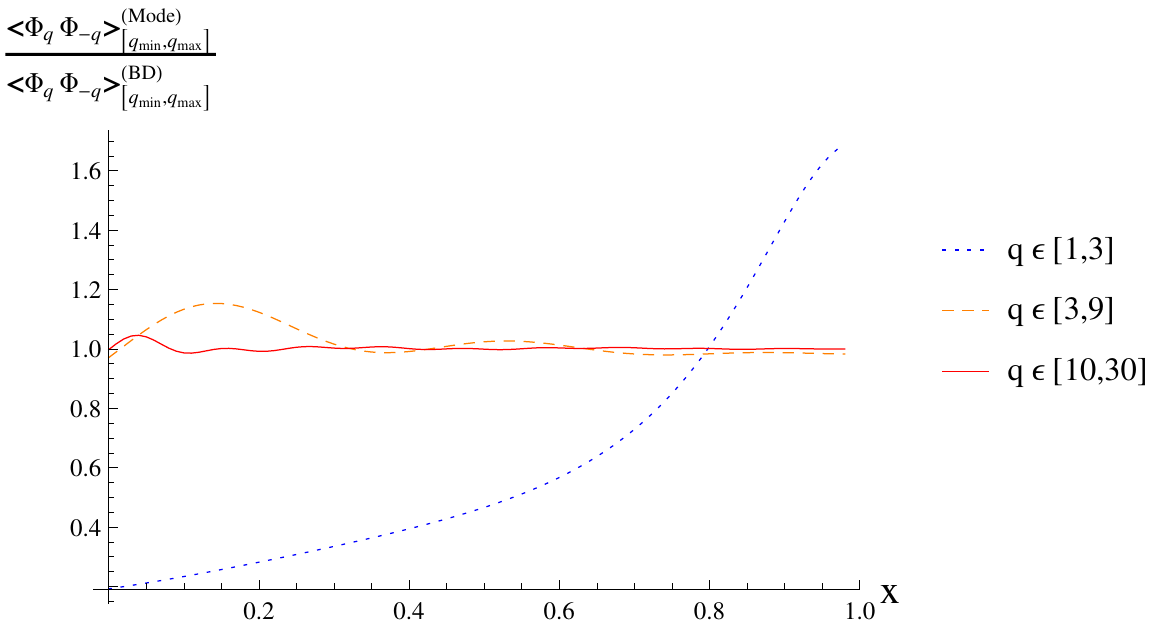}
	\caption{The ratio $\frac{\left\langle\Phi_{\vec{q}}\Phi_{-\vec{q}} \right\rangle^{(Mode)}_{[q_{min}, q_{max}]}}{\left\langle\Phi_{\vec{q}}\Phi_{-\vec{q}} \right\rangle^{(BD)}_{[q_{min}, q_{max}]}}$ integrated for $q \in [1, 3]$, $q \in [3, 9]$, and $q \in [10, 30]$. For $q_{min} = 1$ and $q_{max} = 3$ (dotted line) the ratio clearly deviates from $1$, corroborating the fact that modes near horizon-exit and beyond, do not equilibrate. For other domains $[q_{min}, q_{max}]$ (dashed and solid lines) $\frac{\left\langle\Phi_{\vec{q}}\Phi_{-\vec{q}} \right\rangle^{(Mode)}_{[q_{min}, q_{max}]}}{\left\langle\Phi_{\vec{q}}\Phi_{-\vec{q}} \right\rangle^{(BD)}_{[q_{min}, q_{max}]}}$ oscillates about $1$ with a clear damping over time.}
	\label{phiRatioInt}
\end{figure}

As shown in Fig. \ref{phiRatioInt}, integrating
$\left\langle\Phi_{\vec{q}}\Phi_{-\vec{q}}
\right\rangle^{(Mode)}_{[q_{min}, q_{max}]}$ and
$\left\langle\Phi_{\vec{q}}\Phi_{-\vec{q}}
\right\rangle^{(BD)}_{[q_{min}, q_{max}]}$ over $q$ introduces damping
in the ratio of the two for mode ranges well within the horizon, while
averaging over the near-horizon-crossing range ($q \in [ 1, 3]$) 
does not. For the former modes, the ratios oscillate about $1$ but a damping occurs over
time such that the two-point functions eventually asymptote to $1$.
For $q \in [10, 30]$, the ratio clearly becomes $1$, i.e,
equilibration of $\left\langle\Phi_{\vec{q}}\Phi_{-\vec{q}}
\right\rangle^{(Mode)}_{[q_{min}, q_{max}]}$ with
$\left\langle\Phi_{\vec{q}}\Phi_{-\vec{q}}
\right\rangle^{(BD)}_{[q_{min}, q_{max}]}$ is reached. Looking at
higher $q$-modes, we observed that the higher the $q$-domain the
earlier the equilibration, since such modes have smaller
amplitudes. Comparing with Fig. \ref{phiRatio3q}, we can infer that
the damping is due to the integration over $q$-modes. Hence we may
conclude that, from the perspective of the field correlation
functions, equilibrium is attained for sets of modes that start well
inside the horizon, with $q$ of order $10$ and beyond.  

\begin{figure}[htbp]
	\centering
	\includegraphics[width=5 in, keepaspectratio]{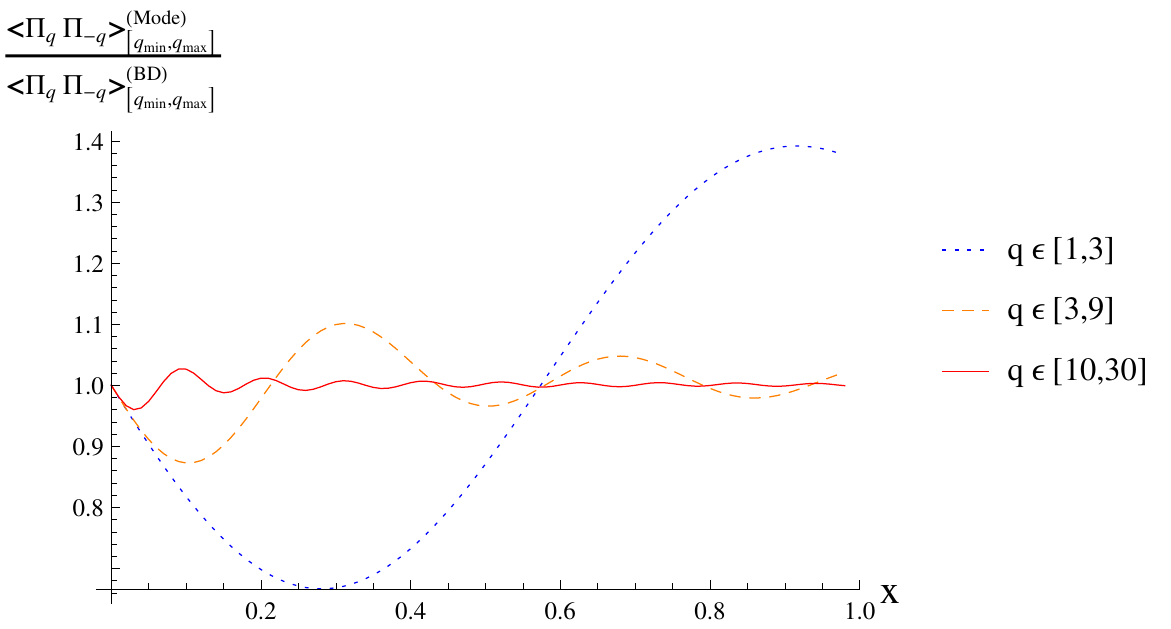}
	\caption{The ratio $\frac{\left\langle\Pi_{\vec{q}}\Pi_{-\vec{q}} \right\rangle^{(Mode)}_{[q_{min}, q_{max}]}}{\left\langle\Pi_{\vec{q}}\Pi_{-\vec{q}} \right\rangle^{(BD)}_{[q_{min}, q_{max}]}}$ integrated for $q \in [1, 3]$, $q \in [3, 9]$, and $q \in [10, 30]$. For $q_{min} = 1$ and $q_{max} = 3$ (dotted line) the ratio oscillates about $1$ without damping, while for other ranges $[q_{min}, q_{max}]$, the ratio damps out close to $1$. For $q \in [10, 30]$ (solid line), it clearly achieves $1$ for higher $x$-values. In other words coarse-graining $\frac{\left\langle\Pi_{\vec{q}}\Pi_{-\vec{q}} \right\rangle^{(Mode)}_{[q_{min}, q_{max}]}}{\left\langle\Pi_{\vec{q}}\Pi_{-\vec{q}} \right\rangle^{(BD)}_{[q_{min}, q_{max}]}}$ over such modes results in equilibration of the numerator and denominator.}
	\label{piRatioInt}
\end{figure}

We may draw similar conclusions from Fig. \ref{piRatioInt}  as in Fig. \ref{phiRatioInt}. For $q \in [3, 9]$ and $q \in [10,30]$, $\frac{\left\langle\Pi_{\vec{q}}\Pi_{-\vec{q}} \right\rangle^{(Mode)}_{[q_{min}, q_{max}]}}{\left\langle\Pi_{\vec{q}}\Pi_{-\vec{q}} \right\rangle^{(BD)}_{[q_{min}, q_{max}]}}$ appears oscillatory about the equilibrium position and damped. For the former domain, the ratio does not exactly achieve equilibrium but approaches it. It clearly hits $1$ for $q \in [10, 30]$. For $q_{min} = 1$, no equilibration occurs. Therefore, we may conclude that modes such that $q$ is of order $10$ and beyond equilibrate, from the point of view of the momentum correlator.  

\begin{figure}[htbp]
	\centering
	\includegraphics[width=5 in, keepaspectratio]{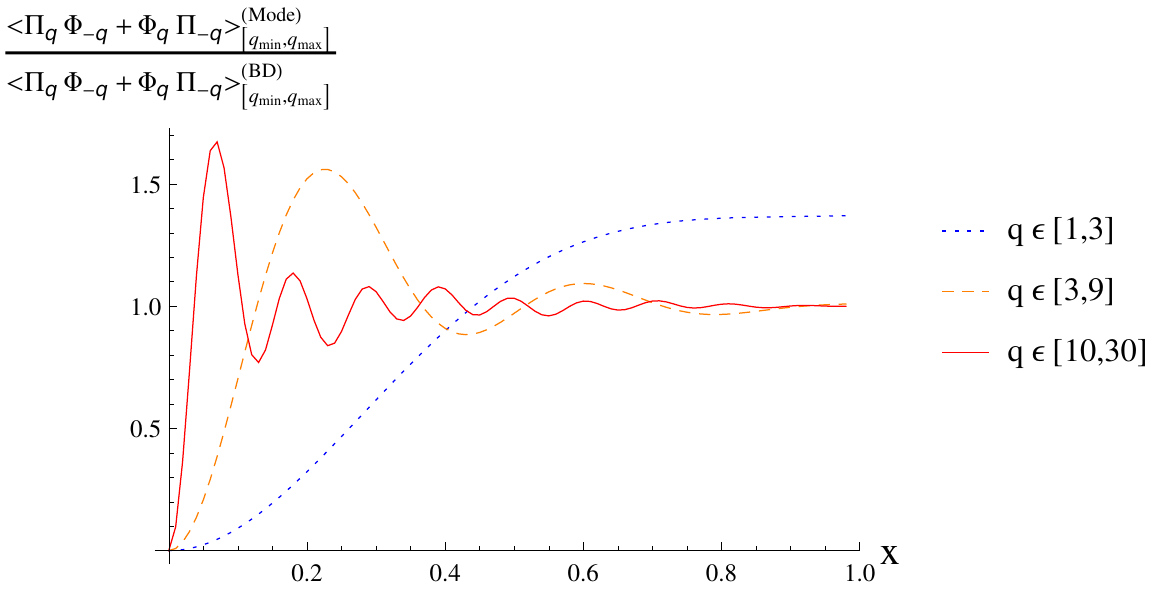}
	\caption{The ratio $\frac{\left\langle \Phi_{\vec{q}}\Pi_{-\vec{q}} +\Pi_{\vec{q}}\Phi_{-\vec{q}} \right\rangle^{(Mode)}_{[q_{min}, q_{max}]}}{ \left\langle \Phi_{\vec{q}}\Pi_{-\vec{q}} +\Pi_{\vec{q}}\Phi_{-\vec{q}} \right\rangle^{(BD)}_{[q_{min}, q_{max}]}}$ integrated for $q \in [1, 3]$, $q \in [3, 9]$, and $q \in [10, 30]$. For the first domain of $q$ (dotted line), $\left\langle \Phi_{\vec{q}}\Pi_{-\vec{q}} +\Pi_{\vec{q}}\Phi_{-\vec{q}} \right\rangle^{(Mode)}_{[q_{min}, q_{max}]}$ never equilibrates to $\left\langle \Phi_{\vec{q}}\Pi_{-\vec{q}} +\Pi_{\vec{q}}\Phi_{-\vec{q}} \right\rangle^{(BD)}_{[q_{min}, q_{max}]}$. For $q_{min} = 3$ or $10$, the ratio is damped over time, and equilibrium is reached for $q$-modes of order and greater than $10$.}
	\label{mixRatioInt}
\end{figure}

In Fig. \ref{mixRatioInt}, we observe a slight difference in the behavior of the lowest $q$-range modes. For $q \in [1, 3]$, $\frac{\left\langle \Phi_{\vec{q}}\Pi_{-\vec{q}} +\Pi_{\vec{q}}\Phi_{-\vec{q}} \right\rangle^{(Mode)}_{[q_{min}, q_{max}]}}{ \left\langle \Phi_{\vec{q}}\Pi_{-\vec{q}} +\Pi_{\vec{q}}\Phi_{-\vec{q}} \right\rangle^{(BD)}_{[q_{min}, q_{max}]}}$ appears to plateau for $x \geq 0.7$. Nevertheless no approach to unity can be found. This again proves that modes which are crossing or near-crossing the horizon do not equilibrate. Plots generated after integrating for $q \in [3, 9]$, and $q \in [10, 30]$, show the same trends as in Fig. \ref{phiRatioInt} and \ref{piRatioInt}. Such modes approach (for $q_{min} = 3$) equilibrium or equilibrate ($q_{min} = 10$ and higher).

In summary, when we ask whether the correlation functions of our state and the BD state approach one other, the answer seems to be that it depends on which modes are being considered. For those that remain well inside the horizon, we see the tendency of our state to approach the BD one, while for low $q$-modes, this does not occur. 

\subsection{Stress-energy tensor}

\begin{figure}[htbp]
	\centering
	\includegraphics[width=5 in, keepaspectratio]{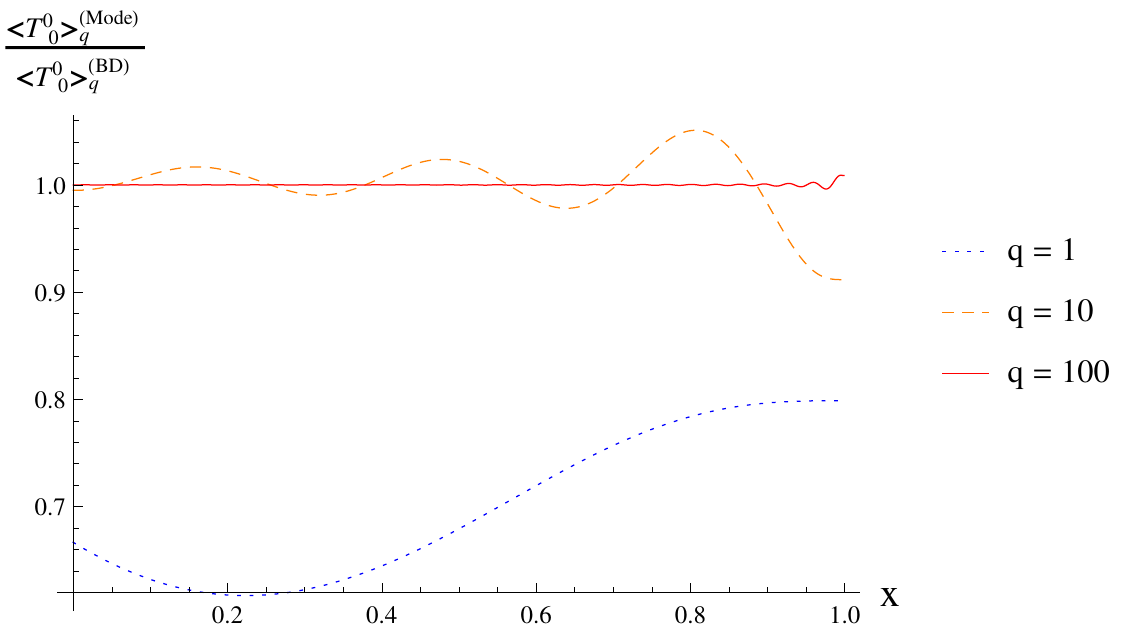}
	\caption{The ratio $\frac{\left\langle T^0_{\hspace{5 pt} 0} \right\rangle_q^{(Mode)}}{\left\langle T^0_{\hspace{5 pt} 0} \right\rangle_q^{(BD)}}$for $q = 1$, $q = 10$ and $q = 100$. For $q = 1$, the ratio settles down well below the equilibrium position, while for $q = 10$ and $100$, it oscillates with increasing amplitude about $1$. None of the curves present equilibration, similar to what was observed for other fine-grained quantities. }
	\label{t00Ratio3q}
\end{figure}

In terms of our variables $x$ and $q$, Eq. \eqref{eq:T00Fourier} in the massless minimally coupled case becomes

\begin{equation}
\label{eq:T00q}
\left\langle T^0_{\phantom{b}0} \right\rangle = \int \frac{d^3q}{(2\pi)^3} \left[ \frac{(1-x)^6}{2} \left\langle\Pi_{\vec{q}}\Pi_{-\vec{q}}\right\rangle + \frac{(1-x)^2}{2}q^2 \left\langle \Phi_{\vec{q}}\Phi_{-\vec{q}}\right\rangle\right]. 
\end{equation}
\\
Let $\left\langle T^0_{\phantom{b}0} \right\rangle_q$ be the integrand of Eq. \eqref{eq:T00q}. Fig. \ref{t00Ratio3q} represents $\frac{\left\langle T^0_{\phantom{b}0} \right\rangle_q^{(Mode)}}{\left\langle T^0_{\phantom{b}0} \right\rangle_q^{(BD)}}$ for $q = 1, 10$ and 100. As seen when analyzing two-point functions, the ratio settles away from $1$ when $q = 1$ and exhibits an oscillatory behavior about 1 for the other $q$-modes. However, contrary to our previous observations, the oscillations are characterized by an amplitude that increases as a function of $x$. This is rather puzzling. Indeed, as discussed in section \ref{subsec:correlators}, our state should be fully described by the two-point functions. $\left\langle T^0_{\phantom{b}0} \right\rangle_q$ itself is a function of two of them, in the massless and minimally coupled case. Thus we should expect to draw the same conclusions as in \ref{subsec:correlators}. Note, however, that our conclusions about equilibration as perceived from the correlators originated after integrating them over $q$. This suggests that we should adopt the same approach here. 

\begin{figure}[htbp]
	\centering
	\includegraphics[width=4 in, keepaspectratio]{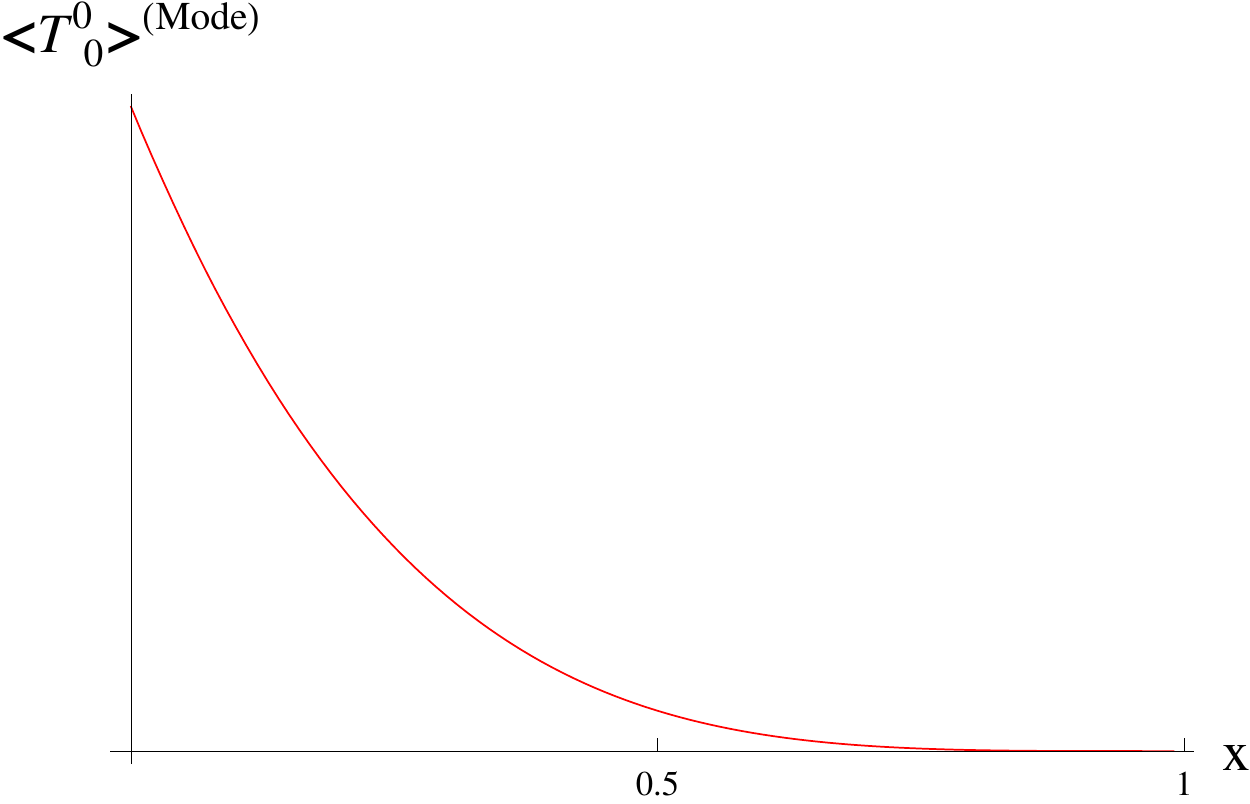}
	\caption{The stress-energy tensor $\left\langle T^0_{\hspace{5 pt} 0} \right\rangle^{(Mode)}_{[q_{min}, q_{max}]}$ corresponding to  $\left\langle T^0_{\hspace{5 pt} 0} \right\rangle_q^{(Mode)}$ integrated between $q = 1$ and $q = 50$. The expectation value monotonically approaches $0$ as $x$ increases.}
	\label{t00Int}
\end{figure}

In Fig. \ref{t00Int} one can observe $\left\langle T^0_{\phantom{b}0} \right\rangle_q^{(Mode)}$ integrated between $q = 1$ and $q = 50$. Let us label it $\left\langle T^0_{\phantom{b}0} \right\rangle^{(Mode)}_{[q_{min}, q_{max}]}$. The distinctive feature of the plot is the fact that the integrated expectation value of the stress-energy tensor decreases monotonically as a function of $x$, and eventually reaches $0$. Given that $\left\langle\Phi_{\vec{q}}\Phi_{-\vec{q}}\right\rangle$ falls off as a function of $x$, and so does $1-x$ (representing the scale factor in our rescaled equations) for $x \in [0,1)$, such a behavior makes sense from the point of view of those quantities. The correlator $\left\langle\Pi_{\vec{q}}\Pi_{-\vec{q}}\right\rangle$, however, rises as a function of $x$. Since it does so as $(1-x)^{-2}$, the presence of $(1-x)^{6}$ in front of $\left\langle\Pi_{\vec{q}}\Pi_{-\vec{q}}\right\rangle$ in Eq. \eqref{eq:T00q} is responsible for an overall decrease. In other words, the expansion of the universe takes care of any diverging behavior in  $\left\langle\Pi_{\vec{q}}\Pi_{-\vec{q}}\right\rangle$.  Looking at $\left\langle T^0_{\phantom{b}0} \right\rangle_q$ $q$-mode per $q$-mode the same declining trend was found, regardless of the chosen vacuum state. Equilibrium however was so far considered from the point of view of ratios of functions evaluated in our state to those evaluated in the Bunch-Davies state. 

\begin{figure}[htbp]
	\centering
	\includegraphics[width=4 in, keepaspectratio]{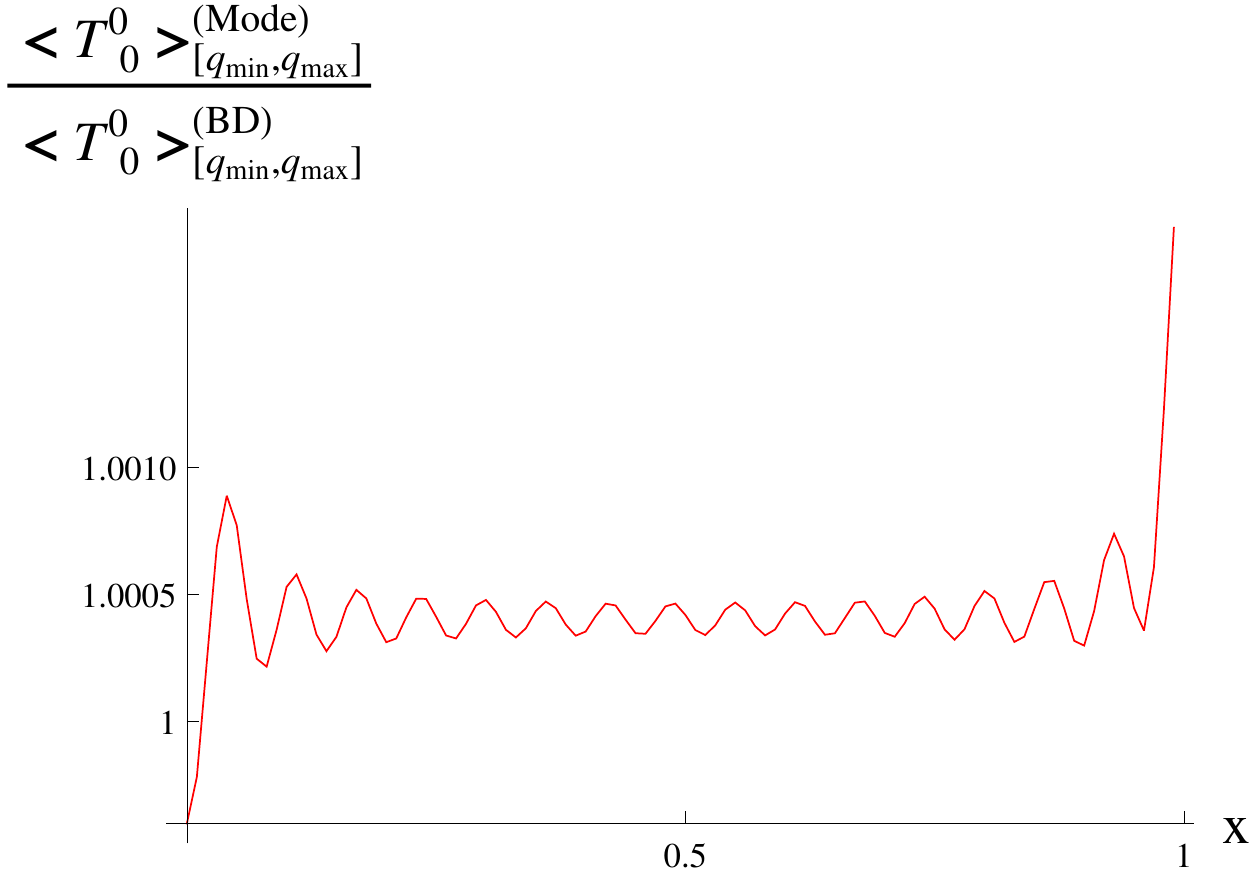}
	\caption{The ratio $\frac{\left\langle T^0_{\hspace{5 pt} 0} \right\rangle^{(Mode)}_{[q_{min}, q_{max}]}}{\left\langle T^0_{\hspace{5 pt} 0} \right\rangle^{(BD)}_{[q_{min}, q_{max}]}}$ obtained using $q_{min} = 1$ and $q_{max} = 50$ as our limits of integration. The ratio seems to oscillate about $1.0004$ with an amplitude that decreases up to $x = 0.5$ but keeps increasing afterward.}
	\label{t00RatioInt}
\end{figure}

The quantity $\frac{\left\langle T^0_{\phantom{b}0} \right\rangle^{(Mode)}_{[q_{min}, q_{max}]}}{\left\langle T^0_{\phantom{b}0} \right\rangle^{(BD)}_{[q_{min}, q_{max}]}}$ is plotted in Fig. \ref{t00RatioInt}. Again, the limits of integration were chosen to be $q_{min} = 1$ and $q_{max} = 50$. The ratio appears to oscillate with damping until about $x = 0.5$, but the amplitude of oscillations keeps increasing afterwards until very late times. This is quite unexpected as, from the point of view of the two-point functions, the ratios appeared damped monotonically with increasing $x$ values, after coarse-graining. 

One could now ask why we used such low limits of integration. Using higher limits resulted in jaggedness in the plots coming from the higher frequency modes in the integral, which are difficult to integrate numerically. Since we are dealing with Hankel functions, themselves highly oscillatory, it is not astonishing that numerical integrators will have difficulty handling them. The fact that a lower step size in $x$ modified the observed jaggedness backs this hypothesis up. 

Smaller steps in $x$ however means greater computing time. Analyzing plots with $q_{max} > 50$ revealed that the equilibrium position of oscillations in $\frac{\left\langle T^0_{\phantom{b}0} \right\rangle^{(Mode)}_{[q_{min}, q_{max}]}}{\left\langle T^0_{\phantom{b}0} \right\rangle^{(BD)}_{[q_{min}, q_{max}]}}$ would decrease to become closer  to $1$. Given the problems encountered after integrating numerically, we chose to call upon Riemann sums of $\left\langle T^0_{\phantom{b}0} \right\rangle_q$ instead of integrals. A step size in $q$ of order unity seemed appropriate and sufficient to draw our conclusions.

\begin{figure}[htbp]
	\centering
	\includegraphics[width=5 in, keepaspectratio]{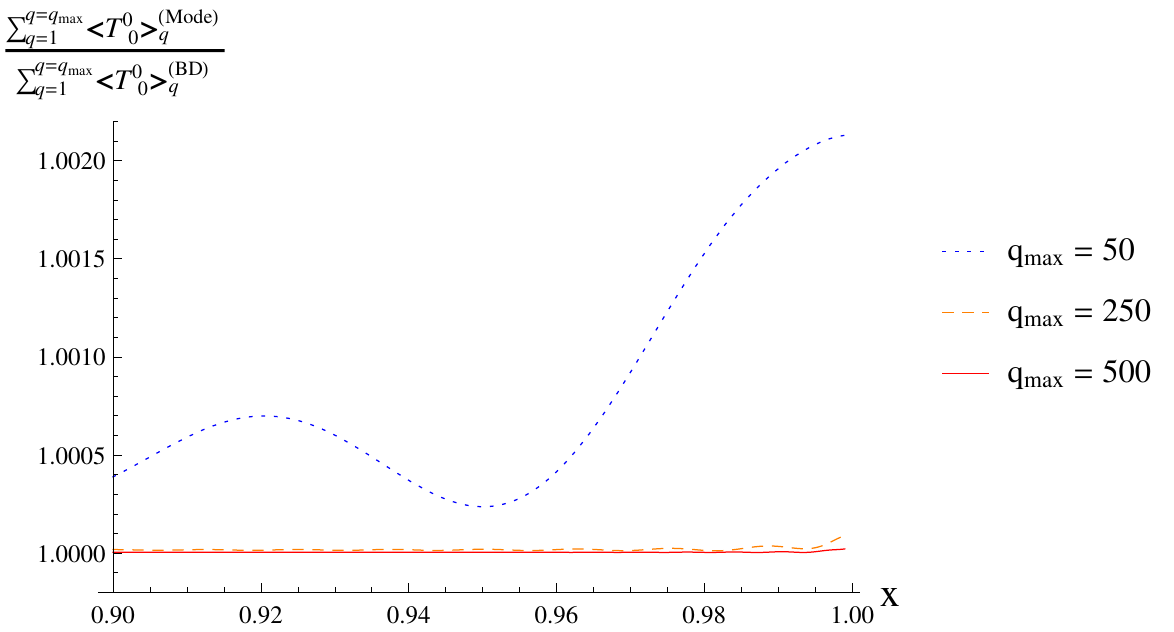}
	\caption{The quantity $\frac{\sum\limits_{q=1}^{q_{max}}\left\langle T^0_{\hspace{5 pt} 0} \right\rangle_q^{(Mode)}}{\sum\limits_{q=1}^{q_{max}}\left\langle T^0_{\hspace{5 pt} 0} \right\rangle_q^{(BD)}}$ for $q_{max} = 50$,  250, and 500, for $x \in [0.900, 0.999]$. Higher $q_{max}$ values correspond to lower amplitude of oscillations of the quotient, and an equilibrium position closer to 1. For $q_{max} = 500$, the ratio is indistinguishable from 1 up to four decimal places.}
	\label{t00SumMaxVar}
\end{figure}

Fig. \ref{t00SumMaxVar} shows $\frac{\sum\limits_{q=1}^{q_{max}}\left\langle T^0_{\phantom{b}0} \right\rangle_q^{(Mode)}}{\sum\limits_{q=1}^{q_{max}}\left\langle T^0_{\phantom{b}0} \right\rangle_q^{(BD)}}$ for $q_{max} = 50,$ 250, and 500 focusing on late times ($x  \in [0.900, 0.999]$). From the plots one can infer that the higher the $q_{max}$ the lower the amplitude of oscillations of our ratio. Additionally, the equilibrium position of the latter gets arbitrarily near 1 as $q_{max}$ increases. For a value $q_{max} =500$, the ratio appears to remain at 1, up to 5 digits. This corroborates the fact that some equilibration process occurs as long as $q$-modes of order 100 and more are included. One last important characteristic that can be found in the figure, is the fact that the quotient does not seem to diverge at very late times but flattens out. 

\begin{figure}[htbp]
	\centering
	\includegraphics[width=5 in, keepaspectratio]{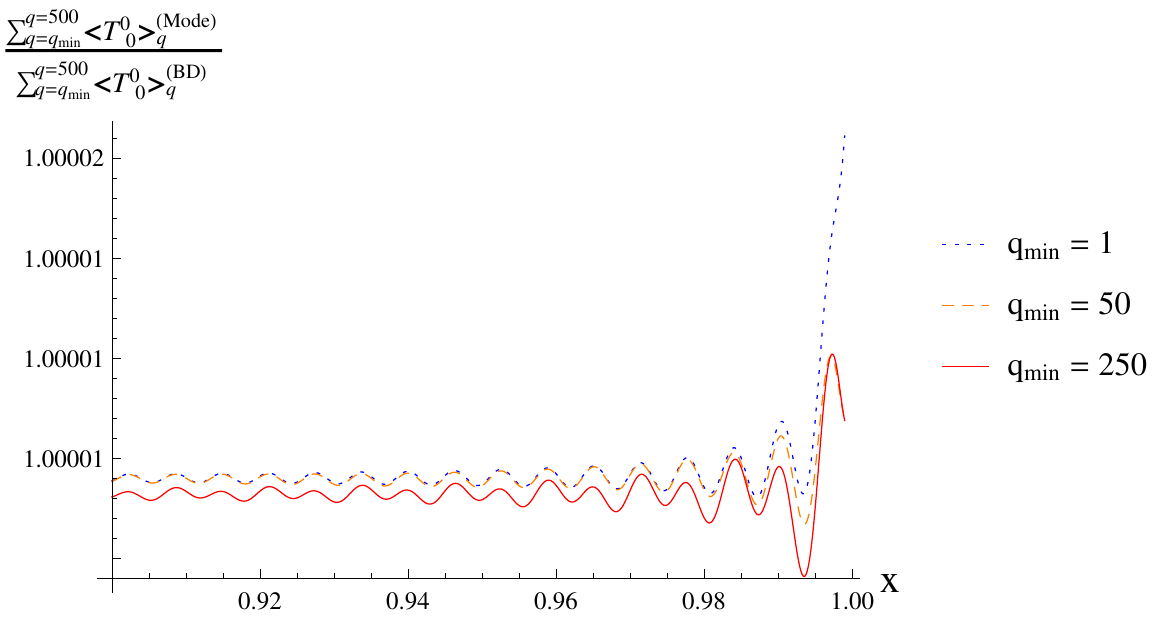}
	\caption{The ratio $\frac{\sum\limits_{q=q_{min}}^{500}\left\langle T^0_{\hspace{5 pt} 0} \right\rangle_q^{(Mode)}}{\sum\limits_{q=q_{min}}^{500}\left\langle T^0_{\hspace{5 pt} 0} \right\rangle_q^{(BD)}}$ for $q_{min} = 1$, 50, and 250, for $x \in [0.900, 0.999]$. The equilibrium position of the quotients and their amplitude of oscillations go down as $q_{min}$ rises.}	
	\label{t00SumMinVar}
\end{figure}

Another path to consider is changing our lower limit $q_{min}$ in the integration, $q_{min} = 1$ corresponding to modes exiting the horizon. As seen in \ref{subsec:correlators} correlators characterizing such modes behave much differently than those for higher $q$ values. Fig. \ref{t00SumMinVar} shows $\frac{\sum\limits_{q=q_{min}}^{500}\left\langle T^0_{\phantom{b}0} \right\rangle_q^{(Mode)}}{\sum\limits_{q=q_{min}}^{500}\left\langle T^0_{\phantom{b}0} \right\rangle_q^{(BD)}}$ for $q_{min} = 1$, 50, and 250. Similar conclusions as the ones obtained in Fig. \ref{t00SumMaxVar} can be drawn (except from the perspective of $q_{min}$), namely the higher the value of $q_{min}$, the closer the central value about which oscillations occur is to unity. Also, the amplitude decreases with higher $q_{min}$'s.

The stress-energy tensor seemed at first to be telling us a slightly different story about the equilibration of our state. However, focusing on the late time behavior of $\frac{\left\langle T^0_{\phantom{b}0} \right\rangle^{(Mode)}_{[q_{min}, q_{max}]}}{\left\langle T^0_{\phantom{b}0} \right\rangle^{(BD)}_{[q_{min}, q_{max}]}}$,  and/or changing the domain of $q$-values to sum over, allows us to reconcile the conclusions coming from our different quantities. Indeed, lower $q$-modes (of order unity) cross the horizon at earlier values of $x$ and have a much larger amplitude of oscillations as compared to the modes with $q$-values that are one or more orders of magnitude higher.  

Our relatively low values of $q_{max}$ means any disparity from equilibrium would become mostly washed out by increasing $q_{max}$ by a factor of 10 and using proper integration techniques. Nevertheless one should still expect the late time increase to be observed with increasing precision, even though it would decrease in amplitude. A reasonable explanation lies in the horizon-exit of the modes at different times. Such modes are still accounted for in our sums at late times, responsible for the seemingly anomalous rise of $\frac{\left\langle T^0_{\phantom{b}0} \right\rangle^{(Mode)}_{[q_{min}, q_{max}]}}{\left\langle T^0_{\phantom{b}0} \right\rangle^{(BD)}_{[q_{min}, q_{max}]}}$. Hence, the effects observed in Fig. \ref{t00RatioInt} should be attributed to the limit in precision, the bounds in $q$ of our numerical calculation, and horizon-crossing effects.

\section{Conclusions}

To check for the approach to equilibrium for any given system, said system has to be disturbed from the putative equilibrium state. Then, its relaxation, or lack thereof back to the original state can be studied. This is what we did here: we disturbed de Sitter space by attaching to it a flat space segment for conformal times $\eta \leq \eta_0$, and considered what happened to the quantum state of a test scalar field in this geometry. The claim being tested is that this state should relax to the ``thermal'' Bunch-Davies state. 

The simplicity of the system allowed us to fully characterize the state by its various two-point functions and we used the ratios of these two-point functions in the disturbed state to their values in the BD state as our diagnostics of equilibration. We also used the stress-energy tensor as a check on whether our state evolved to the BD one. 

What we found was that if we considered these quantities mode by mode there was no evidence of equilibration.  Coarse-grained quantities (integrated over a range of $q$-modes) did show evidence of equilibration in cases where the modes were well within the horizon.  It seems that the importance of coarse-graining in our analysis here is no different than it is in more familiar equilibrating systems.   Integrating our quantities over momentum modes that remained inside a horizon defined by $\eta_0$, we saw that they did equilibrate, and the smaller the wavelengths of our modes, the earlier the equilibration. Particular attention was given to the analysis of the momentum-energy tensor which, initially, presented disparities when compared to the other observables. The range of horizon-exit times of our modes, the necessity to restrict our domain in $q$-space, and a finite precision in our numerical integrations accounted for such differences.  

While our results corroborate the attractor behavior of the Bunch-Davies state, no notion of thermality was discussed. The Bunch-Davies state however is considered thermal \cite{BirrellDavies1982}, upon calculation of how an Unruh-DeWitt detector responds when the field is placed in such a state. Such ambiguity will be investigated and hopefully alleviated in future work (\cite{AlbrechtRichardHolman2014}) in which we address the thermality of our state, after developing a different way to calculate the response rate of an Unruh-DeWitt detector.

\begin{acknowledgments}
We thank McCullen Sandora for useful discussions. R.~H. was supported in part by the Department of Energy under grant DE-FG03-91-ER40682 as well as the John Templeton Foundation. He also thanks the Department of Physics at UC Davis for hospitality while this work was in progress. A.~A. and B.~R. were supported in part by DOE Grants DE-FG02-91ER40674 and DE-FG03- 91ER40674 and the National Science Foundation under Grant No. PHY11-25915. 
\end{acknowledgments}

%\bibliographystyle{plain}
\bibliography{FeelingDeSitterV009PRD}

\end{document}